\documentclass[12pt,thmsa,arabtex]{article}

\usepackage{amssymb}

\usepackage{graphicx}

\input{tcilatex}

\begin{document}

\begin{center}
\textbf{Comment on "Detuning effects in the one-photon mazer" }

~

 Mahmoud Abdel-Aty\footnote{E-mail: abdelatyquant@yahoo.co.uk}

~

{\small Mathematics Department, Faculty of Science, South Valley
University, 82524 Sohag, Egypt,}

\end{center}

{\small In a recent work, Bastin and Martin (B-M) [Phys. Rev. A {\bf
67}, 053804 (2003)] have analyzed the quantum theory of the mazer in
the off-resonant case. However, our analysis of this case refutes
their claim by showing that their evaluation of the coupled
equations for the off-resonant case is not satisfactory. The correct
expression can be obtained by applying an appropriate formulae for
the involved dressed-state parameters.}


{\bf PACS:} 42.50k, 32.80.-t, 03.65.Ge

The tremendous progress of ultra-cold atomic physics in the past
decades is mainly due to the invention of powerful techniques for
trapping and cooling atoms. The coupling between the center-of-mass
motion of the atom and localized laser fields is a central topic in
quantum optics, and leads to many important effects and applications
in cooling, trapping, deflection, and isotope separation
experiments. The interaction of ultracold atoms with microwave
cavities has been considered in Refs. [1]. These studies treated the
interaction between an incident atom in an excited state and a
cavity field, taking the quantum mechanical center-of-mass motion of
the atom into account. This interaction leads to a new kind of
induced emission named the mazer action [1].

In a recent work, Bastin and Martin [2] have considered the
interaction of cold atoms with microwave high-Q cavity and have
argued that they have removed the restriction that considered in the
previous work by considering the off-resonant case for two-level
atoms. However, the two-level atoms were assumed to interact with a
single mode of the cavity and the off-resonance case was considered
in Ref. [3,4].

Bastin and Martin [2] have considered the Hamiltonian
\begin{equation}
\hat H=\frac{\hat p^2}{2m}+\omega_o\hat\sigma^{+}\hat\sigma+\omega
\hat a^{\dagger}\hat a+gu(z)\{\hat \sigma\hat a^{\dagger}+\hat
a\hat\sigma^{+}\}.
\end{equation}
Let us write equation (1) in the following form
\begin{eqnarray}
\hat H&=&\frac{\hat p^2}{2m}+\hat V \nonumber
\\
\hat V&=&\frac{\Delta}{2}\hat\sigma_z+\omega(\hat a^{\dagger}\hat
a+\frac12\hat\sigma_z)+gu(z)\{\hat \sigma\hat a^{\dagger}+\hat
a\hat\sigma^{+}\}.
\end{eqnarray}
It is easy to show that in the $2\times 2$ atomic-photon space the
eigenvalues and eigenfunctions of the interaction Hamiltonian $\hat
V$
\begin{equation}
\hat V|\Phi_{n}^{\pm}\rangle=E_{n}^{\pm}\Phi_{n}^{\pm}\rangle,
\end{equation}
\begin{eqnarray}
E_{n}^{\pm}&=&(n+\frac12)\omega\pm\sqrt{\frac{\Delta^2}{4}+g^2u^2(z)(n+1)},
\\
|\Phi_{n}^{+}\rangle&=&\cos\theta_n|n+1,g\rangle+\sin\theta_n|n,e\rangle,
\nonumber
\\
|\Phi_{n}^{-}\rangle&=&-\sin\theta_n|n+1,g\rangle+\cos\theta_n|n,e\rangle.
\end{eqnarray}
In the atom-field coupling inside the cavity is considered to be a
constant along the propagation axis of the atoms, the problem can be
solved analytically. In this case, the mesa mode function is given
by $u(z)=1 $ for $0<z<L,$ where $L$ is the length of the cavity in
the z-direction.

Now let us look more carefully at the general case, i.e. we go
beyond the mesa mode case. In this case the orthonormal functions
$|\Phi_{n}^{\pm}\rangle$ in the $2\times 2$ system diagonalize the
Hamiltonian and its elements are diagonal in this set of functions
with
\begin{eqnarray}
V_{n}^{\pm}&=&E_{n}^{\pm}
 \nonumber
\\
cot 2\theta_n&=&-\frac{\frac{\Delta}{2}}{gu(z)\sqrt{n+1}}.
\end{eqnarray}
The states $|\Phi_{n}^{\pm}\rangle$ are z-dependent through the
trigonometric functions, they satisfy
\begin{eqnarray}
\frac{\partial}{\partial
z}|\Phi_{n}^{\pm}\rangle&=&\pm|\Phi_{n}^{\mp}\rangle\frac{d\theta_n}{dz},
 \nonumber
\\
\frac{\partial^2}{\partial
z^2}|\Phi_{n}^{\pm}\rangle&=&\pm|\Phi_{n}^{\mp}\rangle\frac{d^2\theta_n}{dz^2}
-|\Phi_{n}^{\pm}\rangle\left(\frac{d\theta_n}{dz}\right)^2.
\end{eqnarray}
Then $|\Psi(z,t)\rangle$ can be expanded in the form
$|\Psi(z,t)\rangle=\sum_nC_n^{\pm}|\Phi_{n}^{\pm}\rangle$ and it
satisfies the Schr\"odinger equation
\begin{eqnarray}
i\frac{\partial}{\partial z}|\Psi(z,t)\rangle=\hat
H|\Psi(z,t)\rangle.
\end{eqnarray}
Hence the coefficients $C_n^{\pm}(z,t)$ satisfy the coupled
equations
\begin{eqnarray}
\frac{\partial C_n^+}{\partial
t}&=&\left(-\frac{1}{2m}\frac{\partial^2}{\partial
z^2}+V_n^+-\left(\frac{d\theta_n}{dz}\right)^2\right)C_n^+
 \nonumber
\\
&&-\left(2\frac{C_n^-}{\partial
z}\left(\frac{d\theta_n}{dz}\right)+C_n^-\left(\frac{d\theta_n}{dz}\right)^2
\right) \nonumber
\\
\frac{\partial C_n^-}{\partial
t}&=&-\left(-\frac{1}{2m}\frac{\partial^2}{\partial
z^2}+V_n^--\left(\frac{d\theta_n}{dz}\right)^2\right)C_n^-
 \nonumber
\\
&&+\left(2\frac{C_n^+}{\partial
z}\left(\frac{d\theta_n}{dz}\right)+C_n^+\left(\frac{d\theta_n}{dz}\right)^2
\right).
\end{eqnarray}
These equations should replace equations (5a) and (5b) of the B-M
paper [2]. But once $u(z)$ is taken to be constant, then
$\frac{d\theta_n}{dz}$ will vanish and we get the results of [3].

Also, it is important to point out that, in the last line of
equation (5a) of B-M paper, $\theta\hbar\delta\sin 2$ should be
replaced by $\hbar\delta\sin 2\theta$.

The most serious point is that Bastin and Martin have overlooked the
formulae for $\cos2\theta_n$ and $\sin2\theta_n$, as well as
\begin{eqnarray}
\cos2\theta_n&=&\frac{-\Delta/2}{\sqrt{\frac{\Delta^2}{4}+g^2u^2(z)(n+1)}},
\nonumber \\
\sin2\theta_n&=&\frac{gu(z)\sqrt{n+1}}{\sqrt{\frac{\Delta^2}{4}+g^2u^2(z)(n+1)}}.
\end{eqnarray}
Once these formulae inserted in the corrected equations (5a) and
(5b) of B-M paper, we find that, the second terms vanish identically
and hence the results of B-M paper reduce to the same equations
which have been presented in Ref. [3,4], in which we have described
in details the off-resonant case on the mazer properties.

{\bf Acknowledgments:} I am grateful to Prof. Abdel-Shafy F. Obada
for very constructive comments and helpful discussions.

\end{document}